\documentclass[12pt]{article} 

\usepackage{mathtools, amsfonts, amsthm, latexsym, amssymb,dsfont}
\usepackage[T1]{fontenc}
\usepackage[tracking=true]{microtype}
\usepackage{lmodern}

\usepackage[margin=1in]{geometry}

\usepackage{graphicx}
\usepackage{subcaption}
\usepackage{booktabs}
\usepackage{longtable}
\usepackage{cite}

\usepackage{color}
\definecolor{darkblue}{cmyk}{0.9,0.9,0,0}
\definecolor{wine-stain}{rgb}{0.5,0,0}
\usepackage[colorlinks, allcolors=darkblue, linktoc=all]{hyperref} 
\begin{document}
\thispagestyle{empty}
\vspace*{1in} 

\begin{center}
{\Large \textbf{Bootstrapping mixed correlators in the 2D Ising model}}
\vspace{.6in}

\textbf{Anton de la Fuente} 
\vspace{.2in} 

\textit{Theoretical Particle Physics Laboratory (LPTP), Institute of Physics, EPFL \\  Lausanne, Switzerland} 
\end{center}
\vspace{.2in}

\begin{abstract}
Bootstrapping mixed correlators in three dimensional conformal field theories with a~$\mathbb Z_2$ global symmetry has previously led to a closed allowed region in~($\Delta_\sigma$, $\Delta_\epsilon$) space surrounding the 3D Ising model. We repeat that analysis in two dimensions. By further assuming a gap in the spin-2 sector, we also find a closed allowed region in ($\Delta_\sigma$, $\Delta_\epsilon$) space surrounding the 2D Ising model.
\end{abstract}
\setcounter{page}{0}

\newpage

\section{Introduction}
In the theory of critical phenomena, all properties of a universality class are described by a fixed point of the renormalization group~(RG) (see e.g.\ \cite{cardy, goldenfeld} for textbook treatments). However, RG theory leaves unexplained a striking observation: there are only very few universality classes. Specifically, the classes are generally distinguished by only three properties: the number of dimensions, the global symmetry group of the Hamiltonian, and the number of relevant operators in the fixed point theory. A priori, the RG allows multiple universality classes with those same three properties, but that is not what is found in nature.  

Recently developed tools in conformal field theory (CFT) have shed some light. Fixed points of the RG are often CFTs \cite{scaleReview}. 
A CFT is so constrained by symmetry that all its correlation functions are completely determined by its set of 2- and 3-point functions, which are in turn determined by a discrete set of numbers, known as the ``conformal data'' of the CFT. In order for the conformal data to consistently determine all higher-point functions, they must satisfy a highly nontrivial set of consistency conditions, known as the ``conformal bootstrap equations'' (see \cite{slavaLectures, DSDlectures, rmpReview} for reviews).  The revival of the bootstrap program \cite{Rattazzi} was due to the discovery that numerical methods can efficiently identify large regions in the space of conformal data that are inconsistent with the conformal bootstrap equations.

Because of this, a CFT explanation for why there are only very few universality classes has been suggested by the work of \cite{mixed1, sdpb, mixed2, mixed3}.  They present evidence that the 3D Ising CFT is the only $\mathbb Z_2$-symmetric 3D CFT with exactly two relevant operators. 
In this note, we present similar evidence in two dimensions.

\section{Bootstrap Constraints}
As in \cite{mixed1, mixed3}, we will make the following assumptions about the conformal data:
\begin{enumerate}
	\item The CFT has a $\mathbb{Z}_2$ global symmetry.
	\item There is exactly one relevant $\mathbb Z_2$-odd operator (denoted by $\sigma$ with dimension $\Delta_\sigma$) and one relevant $\mathbb Z_2$-even operator (denoted by $\epsilon$ with dimension $\Delta_\epsilon$). 
	\item There exists a stress tensor and its Ward identities hold. More precisely, we impose that the OPE coefficients satisfy $\lambda_{\sigma\sigma T}/\lambda_{\epsilon\epsilon T} = \Delta_\sigma/\Delta_\epsilon$.
	\item There is a gap in the spin-2 sector\footnote{This assumption was first explored in \cite{spin2gap} for the 3D Ising Model using just a single correlator. We repeated that single correlator analysis in 2D and indeed found an island for $\delta \gtrsim 1$ at $\Lambda = 19$. However, this is weaker than what we found using mixed correlators, where $\delta \gtrsim 0.8$ is enough to obtain an island.}. More precisely, the next spin-2 operator $T'$ has dimension $\Delta_{T'}= 2+\delta$.
	\item The CFT is unitary. Aside from the restrictions imposed by assumptions 2 and 4, all other operators consistent with unitarity are allowed.
\end{enumerate}
In order to exclude the possibility of having multiple operators with dimensions $\Delta_\sigma$ and $\Delta_\epsilon$, we also need to fix the ratio $\lambda_{\epsilon\epsilon\epsilon}/\lambda_{\sigma\sigma\epsilon}$ (see \cite{mixed1, mixed3} for more details). 

All together, this gives us 4 parameters:
\[
	\{\Delta_\sigma,\Delta_\epsilon, \lambda_{\epsilon\epsilon\epsilon}/\lambda_{\sigma\sigma\epsilon}, \delta \}.
\]
We scanned over these 4 parameters and determined which regions in this parameter space are ruled out by the conformal bootstrap equations.

\section{Results}
Our results are displayed in Figure~\ref{results}. Details of the numerical implementation are given in Appendix~\ref{appendix}.

For fixed spin-2 gap $\delta$, we computed the allowed region in $\{\Delta_\sigma,\Delta_\epsilon, \lambda_{\epsilon\epsilon\epsilon}/\lambda_{\sigma\sigma\epsilon}\}$ parameter space in a neighborhood surrounding the 2D Ising model, which is located at (1/8, 1).
In Figure~\ref{dimensions}, we plot the projection of this allowed region onto the $(\Delta_\sigma, \Delta_\epsilon)$ plane for $\delta = \{0.85, 0.9, 0.95,1\}$. As $\delta$ decreases, the size of this allowed region increases. At $\delta=0.8$ (not shown) the allowed region becomes so large that we were no longer able to identify a closed region in ($\Delta_\sigma$, $\Delta_\epsilon$) space surrounding the 2D Ising model.

For each allowed point, we also computed bounds on OPE coefficients \cite{mixed3}.
In Figure~\ref{ope}, we plot the bound on the OPE coefficients for each allowed point with $\delta =1$. Note that in the actual 2D Ising model, $\lambda_{\epsilon\epsilon\epsilon}$ vanishes.

\begin{figure}
        \centering
        \begin{subfigure}[t]{0.47\textwidth}
                	\centering
                	\includegraphics[width=\textwidth]{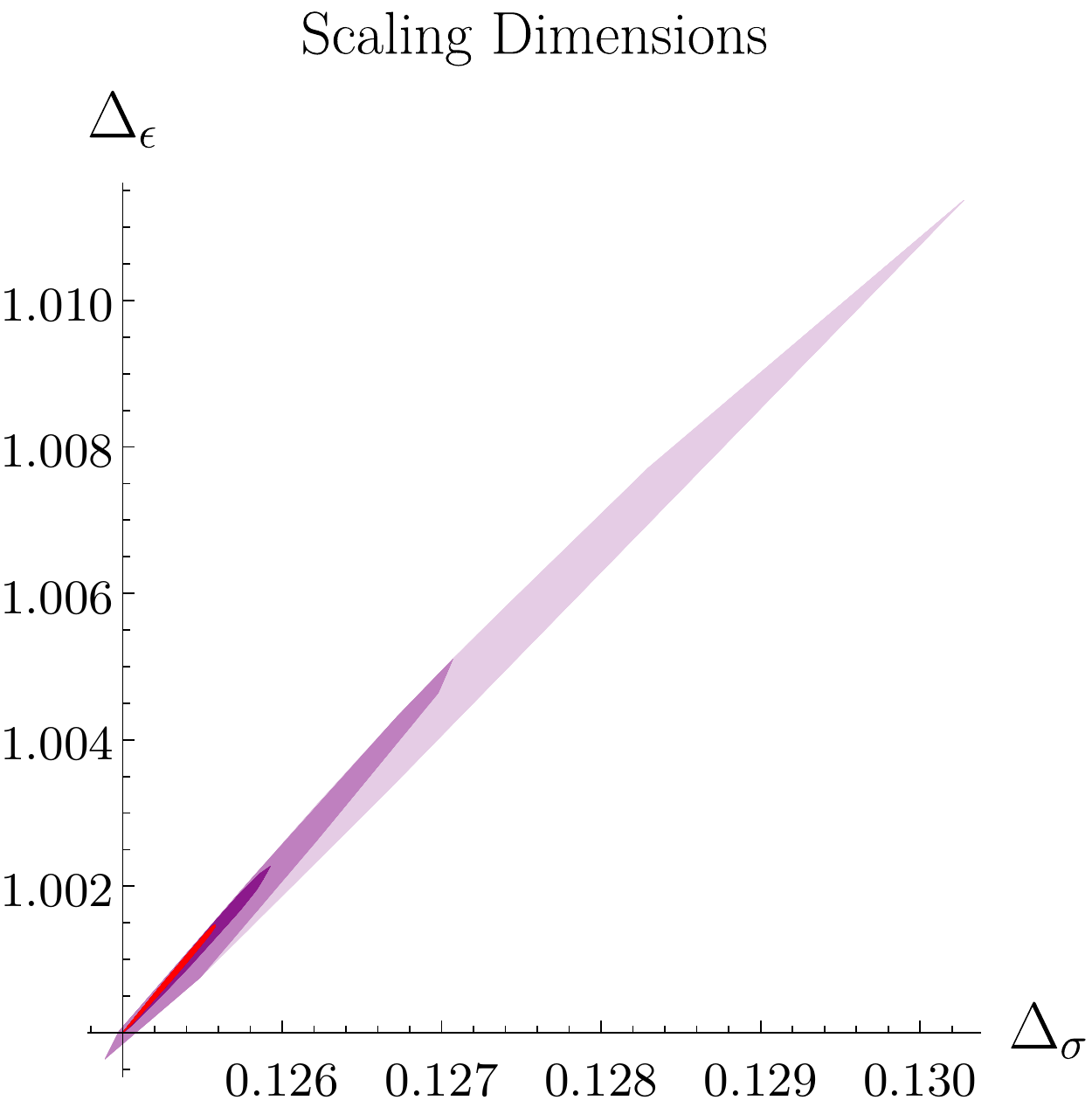}
                	\caption{Allowed region in $\{\Delta_\sigma,\Delta_\epsilon, \lambda_{\epsilon\epsilon\epsilon}/\lambda_{\sigma\sigma\epsilon}\}$ space projected onto the ($\Delta_\sigma$, $\Delta_\epsilon$) plane. Computed with the spin-2 gap $\delta = \{0.85, 0.9, 0.95,1\}$. }
                	\label{dimensions}
        \end{subfigure}
        \hfill
        \begin{subfigure}[t]{0.47\textwidth}
                	\centering
                	\includegraphics[width=\textwidth]{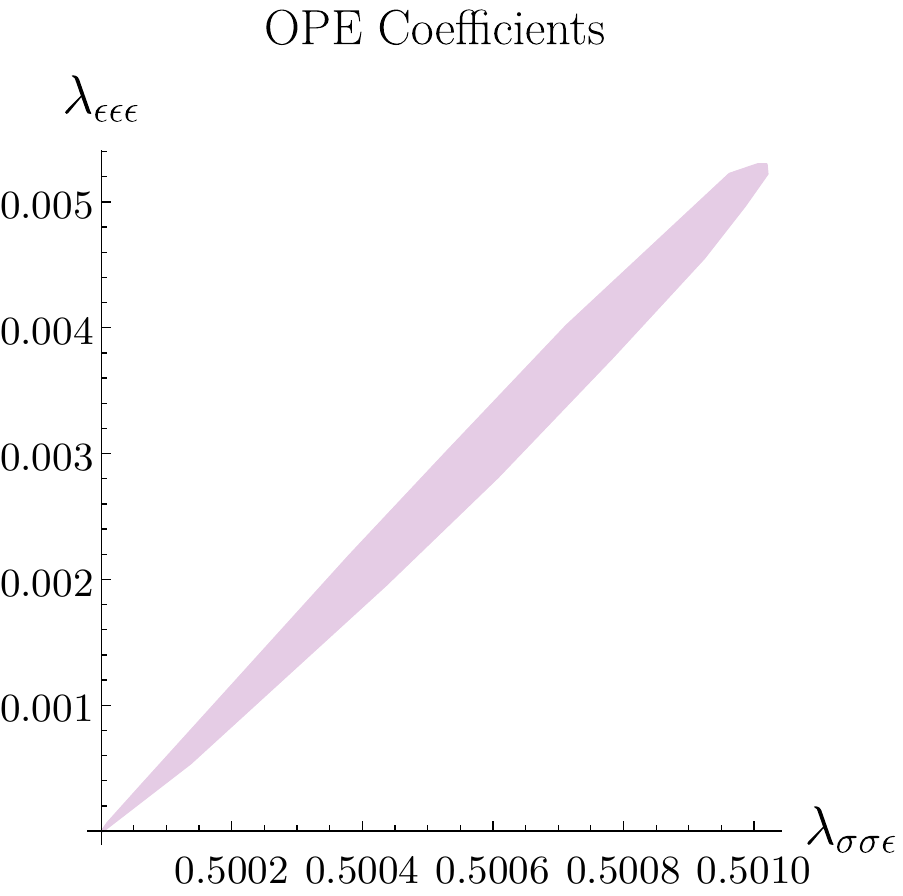}
                	\caption{Computed with $\delta=1$ (red region in Figure~\ref{dimensions}).}
                	\label{ope}
        \end{subfigure}
        \caption{Scaling dimensions and OPE coefficients from the conformal bootstrap.}
        \label{results}
\end{figure}

\section{Discussion}
We presented evidence for the following conjecture: The 2D Ising CFT is the only $\mathbb Z_2$-symmetric 2D CFT with exactly two relevant operators.  More precisely, by assuming that $\sigma$ and $\epsilon$ are the only relevant scalars and that there exists a gap in the spin-2 sector, we found a closed region in $(\Delta_\sigma, \Delta_\epsilon)$ parameter space surrounding the 2D Ising model that is consistent with the conformal bootstrap equations.

Given that the 2D Ising model is exactly solvable \cite{yellowBook}, one might ask whether we found anything new in this work. The answer is yes. Although we did not learn anything new about the 2D Ising model itself, we did learn about the non-existence of other CFTs with two relevant operators. This is true regardless of whether the putative 2D CFT is solvable or not.

There are two straightforward ways to further investigate the above conjecture. The first is to study how the allowed region shrinks as the numerics are improved (see \cite{isingZoom} for a nice illustration in the 3D case). The second is to map out the rest of the allowed region away from the 2D Ising model but with $\Delta_{\sigma}$ and $\Delta_{\epsilon}$ still less than 2.

A natural next step would be to see if the other minimal models can also be isolated by their global symmetry group and number of irrelevant operators \cite{rastelli}.
Discovering already known theories using the conformal bootstrap sharpens it as a tool for carving out the landscape of conformal field theories.

\section*{Acknowledgments}
We thank Connor Behan, Filip Kos, David Poland, David Simmons-Duffin, and Alessandro Vichi for discussions and correspondence. The work of A.D.\ is partially supported by the Swiss National Science Foundation under contract 200020-169696 and through the National Center of Competence in Research SwissMAP. 
\newpage
\appendix
\section{Implementation Details} \label{appendix}
Computations were done with the semidefinite program solver \texttt{SDPB} \cite{sdpb} and closely follows the methods used in \cite{mixed1, sdpb, mixed2, mixed3}. The \texttt{SDPB} parameters used for the computations in this work were:
\begin{center}
\begin{tabular}{lc}
\toprule
\multicolumn{2}{c}{\texttt{SDPB} parameters for scaling dimensions} \\
\midrule
$\Lambda$ & 19 \\
$\kappa$ & 20 \\
\texttt{precision} & 896 \\
\texttt{findPrimalFeasible} & \texttt{True} \\
\texttt{findDualFeasible} & \texttt{True} \\
\texttt{detectPrimalFeasibleJump} & \texttt{True}  \\
\texttt{detectDualFeasibleJump} & \texttt{True}  \\
\texttt{primalErrorThreshold} & $10^{-75}$ \\
\texttt{dualErrorThreshold} & $10^{-30}$ \\
\texttt{initialMatrixScalePrimal} & $10^{40}$ \\
\texttt{initialMatrixScaleDual} & $10^{40}$ \\
\texttt{choleskyStabilizeThreshold} & $10^{-100}$ \\
\texttt{maxComplementarity} & $10^{160}$ \\
\bottomrule
\end{tabular}
\quad
\begin{tabular}{lc}
\toprule
\multicolumn{2}{c}{\texttt{SDPB} parameters for OPE coefficients} \\
\midrule
$\Lambda$ & 15 \\
$\kappa$ & 20 \\
\texttt{precision} & 960 \\
\texttt{dualityGapThreshold} & $10^{-30}$ \\
\texttt{primalErrorThreshold} & $10^{-75}$ \\
\texttt{dualErrorThreshold} & $10^{-75}$ \\
\texttt{initialMatrixScalePrimal} & $10^{60}$ \\
\texttt{initialMatrixScaleDual} & $10^{60}$ \\
\texttt{choleskyStabilizeThreshold} & $10^{-200}$ \\
\texttt{maxComplementarity} & $10^{200}$ \\
\bottomrule
\end{tabular}
\end{center}
The spins included were the set $\{0,1, \dots, 26\} \cup \{49,50\}$. 

We used rational approximations for the conformal blocks \cite{mixed1}. These were computed using the exact formula~\cite{osborn} and then expanding in the $\rho$-variable~\cite{OPE,rho}.

\bibliographystyle{utphys}
\bibliography{references} 
\end{document}